\documentstyle[sprocl,psfig]{article}

\def\Journal#1#2#3#4{{#1} {\bf #2}, #3 (#4)}

\def\NPB{{\em Nucl. Phys.} B}
\def\PLB{{\em Phys. Lett.}  B}
\def\PRL{\em Phys. Rev. Lett.}
\def\PRD{{\em Phys. Rev.} D}
\def\ZPC{{\em Z. Phys.} C}
\def\ZPA{{\em Z. Phys.} A}
\def\NPA{{\em Nucl. Phys.} A}

\def\ra{\rightarrow}

\def\al{\alpha}
\def\be{\begin{equation}}
\def\ee{\end{equation}}
\def\bea{\begin{eqnarray}}
\def\eea{\end{eqnarray}}
\def\rd{{\rm d}}
\def\mev{\,{\rm MeV}}
\def\gev{\,{\rm GeV}}

\begin{document}
\thispagestyle{empty}
\hspace*{9.0cm}                        WU-B 96-21\\
\hspace*{10.0 cm}                       July  1996\\            
\\
\begin{center}
{\Large\bf DIQUARK MODEL PREDICTIONS FOR PHOTON 
INDUCED EXCLUSIVE REACTIONS \footnote
{Invited talk presented at the Workshop on Virtual Compton Scattering, 
Clermont-Ferrand (June 1996)}} \\
\vspace*{1.0 cm}
\end{center}
\begin{center}
{\large P. Kroll}\\
\vspace*{0.5 cm}
Fachbereich Physik, Universit\"{a}t Wuppertal, \\
D-42097 Wuppertal, Germany\\[0.3 cm]
\end{center}
\newpage

\setcounter{page}{1}
\title{DIQUARK MODEL PREDICTIONS FOR PHOTON 
INDUCED EXCLUSIVE REACTIONS }

\author{ P.~Kroll }

\address{Fachbereich Physik, Universit\"{a}t Wuppertal, \\
D-42097 Wuppertal, Germany}

\maketitle\abstracts{
The present status of the diquark model for exclusive reactions at 
moderately large momentum transfer is reviewed. That model is a
variant of the Brodsky-Lepage approach in which diquarks are
considered as quasi-elementary constituents of baryons. Recent
applications of the diquark model, relevant to high energy physics with
electromagnetic probes, are discussed: electromagnetic form factors of
baryons in both the space-like and the time-like region,
photoproduction of mesons, two-photon annihilations into
proton-antiproton pairs as well as real and virtual Compton scattering
on which the main emphasis is laid.}

Exclusive processes at large momentum transfer are described in terms
of hard scatterings among quarks and gluons \cite{lep:80}. In this
so-called hard scattering approach (HSA) a hadronic amplitude
is represented by a convolution of process independent
distribution amplitudes (DA) with hard scattering amplitudes to be
calculated within perturbative QCD. The DAs specify the distribution
of the longitudinal momentum fractions the constituents carry. They
represent Fock state wave functions integrated over transverse
momenta. The convolution manifestly factorizes long (DAs) and short
distance physics (hard scattering). The HSA has two characteristic properties,
the power laws and the helicity sum rule. The first property
says that, at large momentum transfer and large
Mandelstam $s$, the fixed angle cross section of a reaction $AB \ra CD$
behaves as 
\begin{equation}
\label{a1}
\vspace*{-0.2cm}
\rd \sigma /\rd t = f (\theta)\, s^{2-n}
\end{equation}
where n is the minimum number of external particles in the hard 
scattering amplitude. The laws (\ref{a1}) are modified by powers of
$\log{s}$. They also apply to form factors:
a baryon form factor behaves as $1/Q^4$, a meson form factor as $1/Q^2$.
The counting rules are found to be in surprisingly good agreement 
with experimental data. Even at momentum transfers as low as 2 GeV
the data seem to respect the counting rules.\\
The second characteristic property of the HSA is the conservation of
hadronic helicity. For a two-body process the helicity sum rule reads
\begin{equation}
\label{a2}
\vspace*{-0.2cm}
\lambda_A + \lambda_B = \lambda_C + \lambda_D.
\end{equation}
It appears as a consequence of utilizing the 
collinear approximation and of dealing with (almost) massless quarks
which conserve their helicities when interacting with gluons.    
The collinear approximation implies that the relative orbital angular 
momentum between the constituents has a zero component in the
direction of the parent hadron. Hence the helicities of the
constituents sum up to the helicity of their parent hadron.
The helicity sum rule is violated by $20-30 \%$ by many experimental data.
A particular striking example is the Pauli form factor of the proton
which is measured to be large \cite{bos:92}. Its $Q^2$ dependence (see
Fig.~1) is compatible with a higher twist contribution
\begin{figure}[t]
\vspace*{-0.2cm}
\[
\psfig{figure=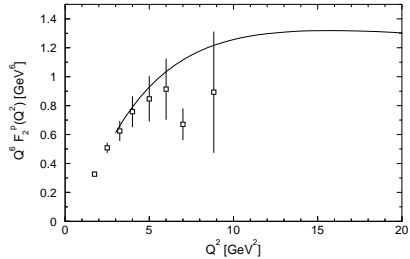,%
        bbllx=0pt,bblly=520pt,bburx=485pt,bbury=830pt,%
        height=3.5cm,clip=}
\]
\vspace*{-0.9cm}
\caption[dummy4]{The Pauli form factor of the proton scaled by
  $Q^6$. Data are taken from \cite{bos:92}. The solid line represents the
  result obtained with the diquark model \cite{jak:93}.}
 \label{fig:pauli}
\vspace*{-0.5cm}
\end{figure}
($\sim 1/Q^6$).\\  
In explicit applications of the HSA (carried through only in leading 
twist and to lowest order QCD with very few exceptions) one encounters
the difficulty that the data are available only at moderately large 
momentum transfer, a region in which non-perturbative dynamics may 
still play a crucial role. A general feature of such applications
is the extreme sensitivity to the DAs chosen for the involved
hadrons. Only strongly end-point concentrated DAs provide results
which are at least for the magnetic form factor of the nucleon in 
fair agreement with the data \cite{cer:84}. This apparent success 
of the HSA is only achieved at the expense of strong contribution 
from soft regions where one of the constituents carries only a tiny 
fraction of its parent hadron's momentum. This is a very problematical situation 
for a perturbative calculation. It should be stressed that none of the
DAs used in actual applications leads to a successful
description of all large momentum transfer processes investigated so far.\\
It seems clear from the above remarks that the HSA at leading twist 
although likely to be the correct asymptotic picture for exclusive 
reactions, needs modifications at moderately large momentum
transfer. In a series of papers \cite{ans:87}$^{-}$\cite{kro:96b} 
such a modification has been proposed
by us in which baryons are viewed as composed of quarks and diquarks. 
The latter are treated as quasi-elementary constituents which partly 
survive medium hard collisions. Diquarks are an 
effective description of correlations in the wave functions and
constitute a particular model for non-perturbative effects.
The diquark model may be viewed as a variant of the HSA appropriate
for moderately large momentum transfer and it is designed in such a way 
that it evolves into the standard pure quark HSA asymptotically. In so
far the standard HSA and the diquark model do not oppose each other, they 
are not alternatives but rather complements.
The existence of diquarks is a hypothesis. However, from experimental
and theoretical approaches there have been many indications suggesting
the presence of diquarks. For instance, they were introduced in baryon
spectroscopy, in nuclear physics, in astrophysics, in jet fragmentation 
and in weak interactions to explain the famous $\Delta I=1/2$ rule. 
Diquarks also provide a natural explanation of the equal slopes of 
meson and baryon Regge trajectories. For more details and for 
references, see \cite{kro:87}. It is important to note that QCD
provides some attraction between two quarks in a colour $\{\bar{3}\}$ 
state at short distances as is to be seen from the static reduction of the
one-gluon exchange term. \\
Even more important for our aim, diquarks have also been found to play a role in
inclusive hard scattering reactions. The most obvious place to signal
their presence is deep inelastic lepton-nucleon scattering. Indeed
the higher twist contributions, convincingly observed \cite{vir:91}, can be
modelled as lepton-diquark elastic scattering. Baryon production in
inclusive $pp$ collisions also reveals the need for diquarks
scattered elastically in the hard interaction \cite{szc:92}. For
instance, kinematical dependences or the excess of
the proton yield over the antiproton yield find simple explanations
in the diquark model. No other explanation of these phenomena is
known as yet.\\
{\bf {\it The diquark model:}} As in the standard HSA a helicity amplitude
for the reaction $AB\ra CD$ is expressed as a convolution of DAs and
hard scattering amplitudes ($s$, $-t$, $-u$\,\, $\gg m_i^2$) 
\begin{eqnarray}
\label{helamp}
{\large M}(s,t)\,=\,\int \rd x_C \rd x_D \rd x_A \rd x_B  
\Phi^*_C(x_C) \Phi^*_D(x_D) T_H(x_i,s,t) \Phi_A(x_A) \Phi_B(x_B) 
\end{eqnarray}
where helicity labels are omitted for convenience. Implicitly
it is assumed in (\ref{helamp}) that the valence Fock states consist
of only two constituents, a quark and a diquarks (antiquark) in the
case of baryons (mesons). In so far the specification of the quark
momentum fraction $x_i$ suffices; the diquark (antiquark) carries the
momentum fraction $1-x_i$. If an external particle is point-like,
e.~g. a photon, the accompanying DA is to be replaced by $\delta (1-x_i)$.
Because of QCD evolution the DAs depend logarithmically on the
momentum transfer. This fact is of minor importance in the limited
range of momentum transfer in which data are available and is
therefore ignored. As in the standard HSA contributions
from higher Fock states are neglected. This is justified by the fact
that that such contributions are suppressed by powers of $\al_s/t$ as
compared to those from the valence Fock state.\\
In the diquark model spin $0$ ($S$) and spin $1$ ($V$) colour
antitriplet diquarks are considered. Within flavour SU(3) the $S$ 
diquarks form an antitriplet, the $V$ diquarks an sixtet. 
Assuming zero relative orbital angular momentum between quark and
diquark and taking advantage of the collinear approximation, the
valence Fock state of an ground state octet baryon $B$ with helicity
$\lambda$ and momentum $p$ can be written in a covariant fashion 
(omitting colour indices)
\begin{equation}
\label{pwf}
|B;p,\lambda\rangle  = f_S\,\Phi_S^B(x)\,B_S\, u(p,\lambda) 
             + f_V\, \Phi_V^B(x)\, B_V
              (\gamma^{\alpha}+p^{\alpha}/m_B)\gamma_5 \,u(p,\lambda)/\sqrt{3}
\end{equation}
where $u$ is the baryon's spinor. The two terms in (\ref{pwf})
represent configurations consisting of a quark and either a scalar or a 
vector diquark, respectively. The couplings of the diquarks 
with the quarks in a baryon lead to flavour functions which e.~g.~for
the proton read
\vspace*{-0.1cm}
\begin{equation}
\label{fwf}
B_S=u\, S_{[u,d]}\hspace{2cm} 
B_V= [ u V_{\{u,d\}} -\sqrt{2} d\, V_{\{u,u\}}]/\sqrt{3}\, .
\end{equation}
The DAs $\Phi^B_{S(V)}$ are conventionally
normalized as $\int \rd x \Phi = 1$. The constants $f_{S(V)}$ play the role the 
configuration space wave function at the origin. \\
The DAs containing the complicated non-perturbative bound state physics, 
cannot reliably be calculated from QCD at present. It is still
necessary to parameterize the DAs and to fit the eventual free
parameters to experimental data. Hence, both the models, the standard
HSA as well as the diquark model, only get a predictive power when a
number of reactions involving the same hadrons are investigated. In
the diquark model the following DAs have been proven to work
satisfactorily well in many applications \cite{jak:93}$^{-}$\cite{kro:96b}:
\begin{eqnarray}
\label{a10}
\Phi^B_S(x)&\hspace{-0.3cm}=&\hspace{-0.3cm}N^B_S x (1-x)^3 
                                 \exp{\left[-b^2 (m^2_q/x + m^2_S/(1-x))\right]}\\
\Phi^B_V(x)&\hspace{-0.3cm}=&\hspace{-0.3cm}N^B_V x (1-x)^3 (1+5.8\,x - 12.5\,x^2)
\exp{\left[-b^2 (m^2_q/x + m^2_V/(1-x))\right]}. \nonumber 
\end{eqnarray}
These DAs are a suitable adaption of a meson DA obtained by
transforming the harmonic oscillator wave function to the
light-cone. The constants $N^B$ are fixed through the
normalization convention (e.~g. for the proton  $N_S^p = 25.97$ 
and $N_V^p = 22.92$). The DAs
exhibit a mild flavour dependence via the exponential which also
guarantees a strong suppression of the end-point regions. The masses
in (\ref{a10}) are constituent masses since they enter through a rest
frame wave function. For $u$ and $d$ quarks we take $350\mev$ and for
the diquarks $580\mev$. Strange quarks and diquarks are assumed to be
$150\mev$ heavier that the non-strange ones. It is to be stressed that
the quark and diquark masses only appear in the DAs (\ref{a10}); in
the hard scattering kinematics they are neglected. The final
results (form factors, amplitudes) depend on the actual mass values mildly.
The transverse size parameter $b$ is fixed from the assumption of a
Gaussian transverse momentum dependence of the full wave function and
the requirement of a value of $600\mev$ for the mean transverse
momentum (actually $b = 0.498\gev^{-1}$). As the
constituent masses the transverse size parameter is not considered as
a free parameter since the final results only depend on it weakly.\\ 
The hard scattering amplitudes $T_H$ determined by short-distance
physics, are calculated from a set of Feyman graphs relevant to a
given process. Diquark-gluon and diquark-photon vertices appear in
these graphs which, following standard prescriptions, are defined as 
\begin{eqnarray}
\label{vert}
\mbox{SgS}:&& i\,g_s t^{a}\,(p_1+p_2)_{\mu} \nonumber\\
\mbox{VgV}:&& -i\,g_{s}t^{a}\, 
\Big\{
 g_{\alpha\beta}(p_1+p_2)_{\mu}
- g_{\beta\mu}\left[(1+\kappa)\,p_2-\kappa\, p_1\right]_{\alpha} \nonumber\\
&& - g_{\mu\alpha} \left[(1+\kappa)\,p_1-\kappa\, p_2\right]_{\beta} 
\Big\} 
\end{eqnarray}
where $g_s=\sqrt{4\pi\alpha_s}$ is the QCD coupling constant.
$\kappa$ is the anomalous magnetic moment of the vector diquark and 
$t^a=\lambda^a/2$ the Gell-Mann colour matrix. For the coupling of 
photons to diquarks one has to replace $g_s t^a$ by $-\sqrt{4\pi\alpha} e_D$ 
where $\alpha$ is the fine structure constant and $e_D$ is the electrical 
charge of the diquark in units of the elementary charge. The couplings 
$DgD$ are supplemented by appropriate contact terms required by 
gauge invariance.\\
The composite nature of the diquarks is taken into 
account by phenomenological vertex functions. Advice for the parameterization 
of the 3-point functions (diquark form factors) is 
obtained from the requirement that asymptotically the diquark 
model evolves into the standard HSA. Interpolating smoothly  between 
the required asymptotic behaviour and the conventional value of 1 at $Q^{2}=0$, 
the diquark form factors are actually parametrized as
\begin{eqnarray}
\label{fs3}
\vspace*{-0.5cm}
F_{S}^{(3)}(Q^{2})=\frac{Q_{S}^{2}}{Q_{S}^{2}+Q^{2}}\,,\qquad
F_{V}^{(3)}(Q^{2})=\left(\frac{Q_{V}^{2}}{Q_{V}^{2}+Q^{2}}\right)^{2}\,.
\end{eqnarray}
The asymptotic behaviour of the diquark form factors and the connection to 
the hard scattering model is discussed in more detail in Ref.~\cite{kro:87,kro:91}.  
In accordance with the required asymptotic behaviour the $n$-point
functions for $n\geq 4$ are parametrized as
\begin{eqnarray}
\label{fsn}
\vspace*{-0.8cm}
F_{S}^{(n)}(Q^{2})=a_{S}F_{S}^{(3)}(Q^{2})\,,\qquad
F_{V}^{(n)}(Q^{2})=
\left(a_{V}\frac{Q_{V}^{2}}{Q_{V}^{2}+Q^{2}}\right)^{n-3}F_{V}^{(3)}(Q^{2}).
\end{eqnarray}
The constants $a_{S,V}$ are strength parameters. Indeed, since the diquarks in 
intermediate states are rather far off-shell one has to consider 
the possibility of diquark excitation and break-up. Both these possibilities 
would likely lead to inelastic reactions. Therefore, we have not to consider 
these possibilities explicitly in our approach but excitation and break-up 
lead to a certain amount of absorption which is taken into account by the 
strength parameters. Admittedly, that recipe is a rather crude
approximation for $n\geq 4$. Since in most cases the contributions from the
n-point functions for $n\geq 4$ only provide small corrections to
the final results that recipe is sufficiently accurate.\\
{\bf {\it Special features of the diquark model:}} The diquark hypothesis
has striking consequences. It reduces the effective number of
constituents inside baryons and, hence, alters the power laws
(\ref{a1}). In elastic baryon-baryon scattering, for instance, the usual power
$s^{-10}$ becomes $s^{-6} F(s)$ where $F$ represents the effect of
diquark form factors. Asymptotically $F$ provides the missing four
powers of $s$. In the kinematical region in which the diquark model 
can be applied ($-t$, $-u \geq 4 \gev^2$), the diquark form factors 
are already active, i.~e.~they supply a substantial $s$ dependence 
and, hence, the effective power of $s$ lies somewhere between 6 and 10.
The hadronic helicity is not conserved in the diquark model at finite
momentum transfer since vector diquarks can flip their helicities when
interacting with gluons. Thus, in contrast to the standard HSA
spin-flip dependent quantities like the Pauli form factor of the 
nucleon can be calculated.\\
{\bf {\it Electromagnetic nucleon form factors:}} This is the simplest
application of the diquark model and the most obvious place to fix the
various parameters of the model. The Dirac and Pauli form factors of
the nucleon are evaluated from the convolution formula (\ref{helamp}) 
with the DAs (\ref{a10}) and the parameters are determined from a best
fit to the data in the space-like region. The following set of parameters
\begin{equation}
\label{c1}
\begin{array}{cccc}
 f_S= 73.85\,\mbox{MeV},& Q_S^2=3.22 \,\mbox{GeV}^2, & a_S=0.15, &  \\
 f_V=127.7\,\mbox{MeV},& Q^2_V=1.50\,\mbox{GeV}^2, & a_V=0.05,&\kappa=1.39\,;
\end{array}
\end{equation}
provides a good fit of the data \cite{jak:93}. $\alpha_s$ is evaluated with 
$\Lambda_{QCD}=200\mev$ and restricted to be smaller than $0.5$. The 
parameters $Q_S$ and $Q_V$, controlling the size of the diquarks,
are in agreement with the higher-twist effects observed in the structure 
functions of deep inelastic lepton-hadron scattering \cite{vir:91} if these 
effects are modelled as lepton-diquark elastic scattering. The Dirac
form factor of the proton is perfectly reproduced. The results for the
Pauli form factor are shown in Fig.~1.
The predictions for the two neutron form factors are also in
agreement with the data. However, more accurate neutron data are needed in the
$Q^2$ region of interest in order to examine the model crucially. 
The nucleon's axial form factor \cite{jak:93} and its electromagnetic
form factors in the time-like
regions \cite{kro:93a} have also been evaluated. Both the results compare
well with data. Even electroexcitation of nucleon resonances has been
investigated \cite{kro:92,bol:94}.\\
{\bf {\it Real Compton scattering (RCS):}} $\gamma p\ra\gamma p$ is the next
reaction to which the diquark model is applied. Since again the 
only hadrons involved are protons RCS can be predicted
in the diquark model; there is no free parameter to be adjusted. Typical
Feynman graphs contributing to that process are shown in Fig.~2. 
\begin{figure}[t]
\vspace*{-0.5cm}
\[
    \psfig{figure=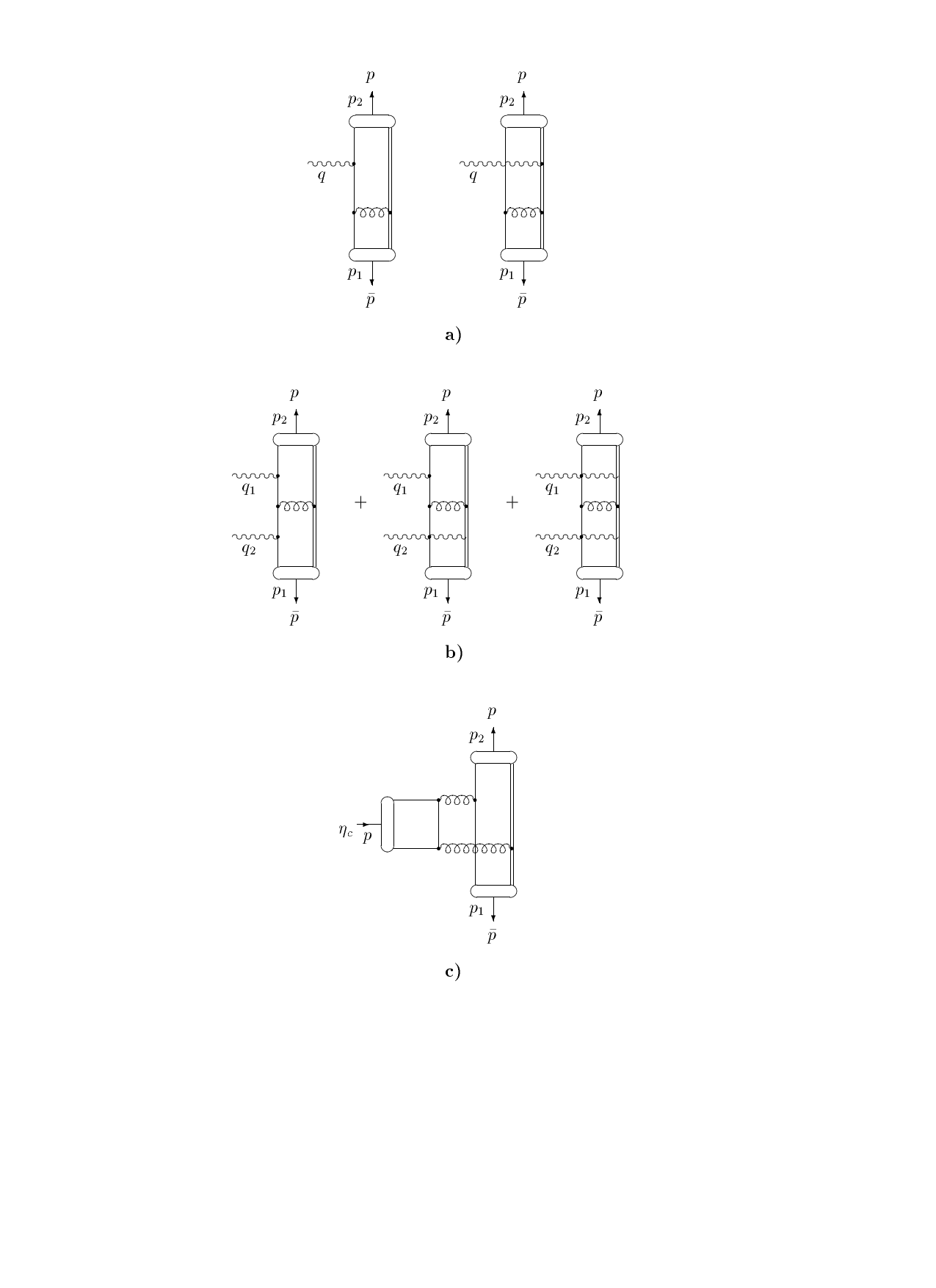,%
        bbllx=140pt,bblly=423pt,bburx=430pt,bbury=600pt,%
        height=3cm,clip=}
\]
\vspace*{-1.0cm}
\caption[]{Typical Feynman graphs contributing to  
$\gamma^{(\ast)}\,p\rightarrow\gamma\,p$.}
\label{frfd}
\end{figure}
The results of the diquark model for RCS are shown
in Fig.~3 for three different photon energies \cite{kro:91,kro:96a}. 
\begin{figure}[t]
\[
    \psfig{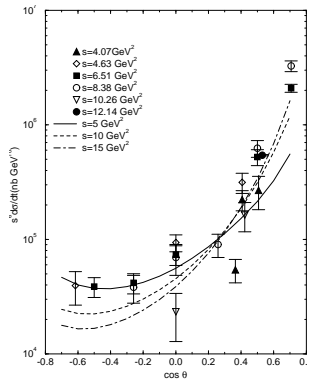}
\]
\vspace*{-1.0cm} 
\caption[]{The cross section for RCS off protons scaled 
by $s^6$ vs.~$\cos\theta$ for three different photon energies. 
The experimental data are taken from \cite{shu:79}.} 
\label{frcc}
\vspace*{-0.5cm} 
\end{figure}
Note that in the very forward and backward regions the transverse
momentum of the outgoing photon is small and, hence, the diquark
model which is based on perturbative QCD, is not applicable. Despite
the rather small energies at which data \cite{shu:79} are available, 
the diquark model is seen to work rather well. The predicted
cross section does not strictly scale with $s^{-6}$. The results obtained within the
standard HSA are of similar quality \cite{niz:91}. The diquark model
also predicts interesting photon asymmetries and spin correlation
parameters (see the discussion in \cite{kro:91}). Even a polarization
of the proton, of the order of $10\%$, is predicted
\cite{kro:91}. This comes about as a consequence of helicity flips 
generated by vector diquarks and of perturbative
phases produced by propagator poles appearing within the domains of
momentum fraction integrations. The poles are handled in
the usual way by the $\imath\varepsilon$ presription. The appearance of imaginary
parts to leading order of $\al_s$ is a non-trivial prediction of
perturbative QCD \cite{far:89}; it is characteristic of the HSA and is not
a consequence of the diquark hypothesis.\\
{\bf {\it Two-photon annihilation into $p\bar{p}$ pairs:}} This
process is related to RCS by crossing, i.~e.~the same set of Feynman
graphs contributes (see Fig.~2). The only difference is that now the 
diquark form factors are needed in the time-like region. The expressions
(\ref{fs3},\ref{fsn}) represent an effective parameterization of them
valid at large space-like $Q^2$. Since the exact dynamics of the
diquark system is not known it is not possible to continue these
parameterizations to the time-like region in a unique way.  
A continuation can be defined as follows \cite{kro:93a}: $Q^2$ is 
replaced by $-s$ in (\ref{fs3},\ref{fsn}) guaranteeing the correct 
asymptotic behaviour and, in order to avoid the appearance
of unphysical poles at low $Q^2$, the diquark form factors are kept
constant once their absolute values have reached $c_0=1.3$ \cite{kro:93a}. 
The same definition of the time-like diquark form
factors is used in the analysis the proton form factor in the
time-like region. The diquark model predictions for the
integrated $\gamma\gamma\ra p\bar{p}$ cross section is compared to the
CLEO data \cite{cleo} in Fig.~4. At large energies the agreement
\begin{figure}[t]
\[
    \psfig{figure=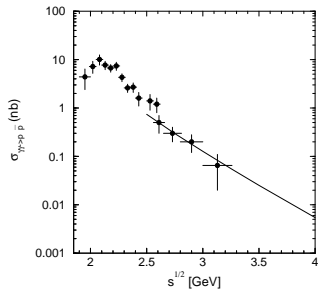,%
        bbllx=70pt,bblly=220pt,bburx=470pt,bbury=573pt,%
        height=4cm,clip=}
\]
\vspace*{-0.9cm}
\caption[]{The integrated $\gamma\gamma\ra p\bar{p}$ cross section
  ($\mid\cos{\theta}\mid\ge 0.6$). The solid line represents the diquark
  model prediction \cite{kro:93a}. Data are taken from CLEO \cite{cleo}.} 
\label{time}
\vspace*{-0.5cm} 
\end{figure}
between predictions and experiment is good. The predictions for the 
angular distributions are in agreement with the CLEO data too. The
standard HSA on the other hand predicts a cross section which lies
about an order of magnitude below the data \cite{far:85}. Recently
CLEO has also measured two-photon annihilations into
$\Lambda\bar{\Lambda}$ pairs \cite{blis:96}. Surprisingly the
integrated cross section is, within errors, as large as that
for annihilations into $p\bar{p}$ pairs. Using the SU(6)-like
spin-flavour dependence (\ref{pwf},\ref{a10}), the diquark model
predicts a $\Lambda\bar{\Lambda}$ cross section which is about a factor
of 2 smaller than the CLEO data. The reason for this discrepancy is
not yet understood. \\
{\bf {\it Virtual Compton scattering (VCS):}} This process is accessible
through $ep \ra ep \gamma$. An interesting element in that reaction 
is that, besides VCS, there is also a
contribution from the Bethe-Heitler (BH) process where the final state 
photon is emitted from the electron. 
Electroproduction of photons offers many possibilities to test
details of the dynamics: One may measure the $s$, $t$
and $Q^2$ dependence as well as that on the angle $\phi$ between the
hadronic and leptonic scattering planes. This allows to
isolate cross sections for longitudinal and transverse virtual
photons. One may also use polarized beams and targets and last but not
least one may measure the interference between the BH
and the VC contributions. The interference is sensitive 
to phase differences.\\ 
At $s$, $-t$ and $-u \gg m_p^2$ (or small $|\cos{\theta}|$ where
$\theta$ is the scattering angle of the outgoing photon in the
photon-proton center of mass frame) the diquark model can also be 
applied to VCS \cite{kro:96a}. Again there is no free parameter in 
that calculation. The relevant Feynman graphs are the same as for RCS
(see Fig.~2). The model can safely be applied for 
$s\ge 10\gev^2$ and $|\cos{\theta}|\le 0.6$. For the future
CEBAF beam energy of $6\gev$ the model is at its limits of applicability. 
However, since the diquark model predictions for real Compton scattering 
do rather well agree with the data even at $s\ge 5\gev^2$ (see Fig.~3)
one may expect similarly good agreement for VCS. Predictions for the VCS 
cross section are shown in Fig.~5. The transverse 
\begin{figure}[t]
\[
    \psfig{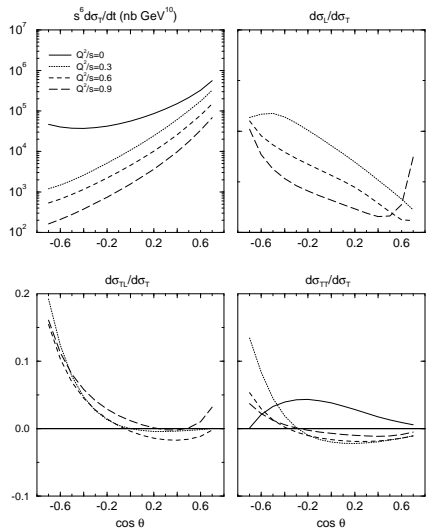}
\]
\vspace*{-1.0cm}
\caption[]{The cross section for VCS vs.~$\cos\theta$ 
for several values of $Q^2/s$ at $s=5\;{\rm GeV}^2$. 
Upper left: the transverse cross section scaled by $s^6$. Upper right: the 
ratio of the longitudinal over the transverse cross sections. Lower left 
(right): the ratio of the longitudinal (transverse) - transverse interference 
term over the transverse cross section.}
\label{fVC5}
\vspace*{-0.5cm}
\end{figure}
cross section (which, at $Q^2=0$, is the cross section for RCS is
the dominant piece. The other cross sections only become sizeable for
large values of $|\cos{\theta}|$. Examination of the Bethe-Heitler
contribution to the process $ep\to ep\gamma$ reveals that it is small
as compared to the VCS contribution at high energies, small values of 
$|\cos{\theta}|$ and for an out-of-plane experiment, i.~e.~$\phi\ge 50^{\circ}$.\\
The last observable I want to discuss is the electron asymmetry in 
$e p \to e p \gamma$:
\begin{equation}
  \label{asy}
A_L\,=\,\frac{\sigma(+) -\sigma(-)}{\sigma(+)+\sigma(+)}  
\end{equation}
where $\pm$ indicates the helicity of the incoming electron. $A_L$
measures the imaginary part of the longitudinal -- transverse 
interference. The longitudinal amplitudes for VCS
turn out to be small in the diquark
model (hence $A_L^{VC}$ is small). However, according to the model, $A_L$ is
large in the region of strong BH contamination (see
Fig.~6). In that region, $A_L$ measures the relative phase
\begin{figure}[t]
\[
    \psfig{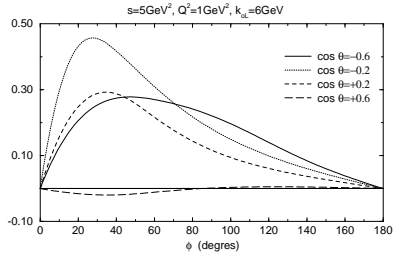}
\]
\vspace*{-1.0cm}
\caption[dummy5]{The electron asymmetry in $ep\to ep\gamma$ as
  predicted by the diquark model \cite{kro:96a}.}
   \label{fig:asym}
\vspace*{-0.5cm}
\end{figure}
(being of perturbative origin from on-shell going internal gluons,
quarks and diquarks \cite{far:89}) between the BH amplitudes
and the VCS ones. The magnitude of the effect shown in
Fig.~6 is sensitive to details of the model and, therefore, should not
be taken literally. Despite of this our results may be taken as an
example of what may happen. The measurement of $A_L$, e.~g.~at CEBAF,
will elucidate strikingly the underlying dynamics of VCS.\\
{\bf {\it Photo- and electroproduction of mesons:}} This is already a quite 
complicated reaction to which all together 158 Feynman graphs
contribute. Up to now only the two processes $\gamma p\ra K^+\Lambda$,
$K^{*+}\Lambda$ have been investigated \cite{kro:96b}. The analyses of
other final states as well as electroproduction  
are in progress. The calculation of $K\Lambda$ production is somewhat
simpler than that for other final states because only scalar diquarks
contribute. The analysis of many different final states will
provide deep insight in the dynamics. \\
As compared to the processes discussed above a new element
appears now, namely the mesonic DA. Comparison of predictions with data 
\cite{and:76} (see Fig.~7) revealed that the asymptotic form of the
Kaon DA ($\sim x(1-x)$) works very well (using the standard value of 
\begin{figure}[t]
\[
    \psfig{figure=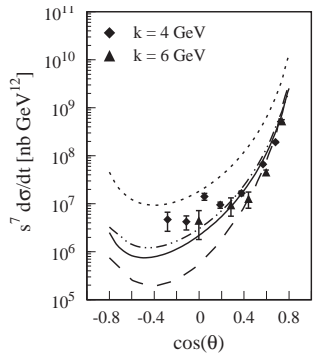,bbllx=30pt,bblly=550pt,bburx=300pt,bbury=810pt,%
                 height=5cm,clip=}
\]
\vspace*{-1.5cm}
\caption[]{Differential cross section for $\gamma p\ra K^+\Lambda$
  scaled by $s^7$ vs.~$\cos{\theta_{cm}}$. Solid (dash-double-dotted)
    line: Diquark model result \cite{kro:96b} at $p^{\gamma}_{lab} = 6
    (4)\gev$ using the asymptotic Kaon DA. Short-dashed line: Results
    $p^{\gamma}_{lab} = 6\gev$ using the CZ Kaon DA. Long-dashed line:
    Predictions from the standard HSA \cite{far:91}. The data are
    taken from Anderson et al. \cite{and:76}.} 
\vspace*{-0.5cm}
\end{figure}
the Kaon decay constant). On the other hand the double-humped DA
proposed by Chernyak and Zhitnitski \cite{cer:84} fails, the predicted
cross section is too large as compared with the data. The predictions 
from the standard HSA \cite{far:91} are smaller than those from the 
diquark model. It should be mentioned that for the $\Lambda$ the 
SU(6)-like spin-flavour dependence (\ref{pwf}, \ref{a10}) is used. How
to reconcile this with the apparent failure of the SU(6)-like
$\Lambda$ wave function in $\gamma\gamma \to \Lambda \bar{\Lambda}$
remains to be seen.  \\
{\it Summary and outlook:} 
The diquark model which represents a variant of the HSA,
combines perturbative QCD with non-perturbative elements. The diquarks
represent quark-quark correlations in baryon wave functions which are
modelled as quasi-elementary constituents. This model has been applied
to many photon induced exclusive processes at moderarely large
momentum transfer (typically $\geq 4 \gev^2$). From the analysis of
the nucleon form factors the parameters specifying the diquark and the
DAs, are fixed. Compton scattering and two-photon annihilations
of $p\bar{p}$ can then be predicted. The comparison with existing
data reveals that the diquark model works quite well and in fact much
better then the pure quark HSA. Using the asymptotic DA for the Kaon
and SU(6) ideas to fix the $\Lambda$ DA one can also predict
photoproduction of $K\Lambda$. Again there is agreement between
predictions and experiment. \\ 
Predictions for the VCS
cross section and for the $ep\to ep\gamma$ cross section have also
been made for kinematical situations accessible at the upgraded CEBAF and 
perhaps at future high energy accelerators like ELFE@HERA. 
According to the diquark model the BH contamination
of the photon electroproduction becomes sizeable for small azimuthal
angles. The BH contribution also offers
the interesting possibility of measuring the relative phases
between the VC and the BH amplitudes. The phases of the VC amplitudes are
a non-trivial phenomenon generated by the fact that some of the internal
quarks, diquarks and gluons may go on mass shell. 
The electron asymmetry $A_L$ is particularly sensitive
to relative phases.
In contrast to the standard HSA the diquark model allows to
calculate helicity flip amplitudes, the helicity sum rule (\ref{a2})
does not hold at finite $Q^2$. One example of an observable controlled
by helicity flip contributions is the Pauli form factor of the
proton. Also in this case the diquark model accounts for the data.

\end{document}